\documentclass[a4paper,11pt]{article}
\usepackage{pos}
\usepackage{float}
\usepackage{subcaption}

\usepackage{lineno}

\newcommand{\esqfig}[5]{
	\begin{figure}
		\centering
		\fbox{\includegraphics[width=#1\textwidth, height=#2\textwidth]{Figures/#3}}
		\caption{{\small #4.}}
		\label{#5}
	\end{figure}
}
\newcommand{\doublefig}[9]{
	\begin{figure*}[h]
		\centering
		\begin{subfigure}[t]{0.49\textwidth}
			\fbox{\includegraphics[width=\textwidth,height=#1\textwidth]{Figures/#2}}
			\caption{{\footnotesize #3.}}
			\label{#4}
		\end{subfigure}
		\begin{subfigure}[t]{0.49\textwidth}
			\fbox{\includegraphics[width=\textwidth,height=#1\textwidth]{Figures/#5}}
			\caption{{\footnotesize #6.}}
			\label{#7}
		\end{subfigure}
		\caption{{\footnotesize #8.}}
		\label{#9}
	\end{figure*}
}

\title{Use of Machine Learning for gamma/hadron separation with HAWC}
 \ShortTitle{Using ML for G/H separation with HAWC}

\author*[a]{T. Capistr\'an}
\author[b]{K. L. Fan}
\author[c]{J. T. Linnemann}
\author[d]{I. Torres}
\author[e,f]{P. M. Saz Parkinson}
\author[g]{Philip L.H. Yu}
\forColl{HAWC}

\affiliation[a]{Instituto de Astronom\'ia, Universidad Nacional Aut\'onoma de M\'exico, Ciudad de Mexico, Mexico}

\affiliation[b]{Department of Physics, University of Maryland, College Park, MD, USA}

\affiliation[c]{Department of Physics and Astronomy, Michigan State University, East Lansing, MI, USA}

\affiliation[d]{Instituto Nacional de Astrof\'isica, \'Optica y Electr\'onica, Luis Enrique Erro 1, Tonantzintla, Puebla 72840, M\'exico}

\affiliation[e]{Santa Cruz Institute for Particle Physics, Department of Physics and Department of Astronomy and Astrophysics, University of California at Santa Cruz, Santa Cruz, CA 95064, USA}

\affiliation[f]{Department of Physics and Laboratory for Space Research, The University of Hong Kong, Pokfulam Road, Hong Kong, China}

\affiliation[g]{Department of Mathematics and Information Technology, The Education University of Hong Kong, 10 Lo Ping Road, Tai Po, New Territories, Hong Kong\\[0.25cm]}

\emailAdd{tcapistran@astro.unam.mx}
\emailAdd{klfan@terpmail.umd.edu}
\emailAdd{linneman@msu.edu}
\emailAdd{ibrahim@inaoep.mx}
\emailAdd{pablosp@hku.hk}
\emailAdd{plhyu@eduhk.hk}

\abstract{Background showers triggered by hadrons represent over 99.9\% of all particles arriving at ground-based gamma-ray observatories. An important stage in the data analysis of these observatories, therefore, is the removal of hadron-triggered showers. Currently, the High-Altitude Water Cherenkov (HAWC) gamma-ray observatory employs an algorithm based on a single cut in two variables, unlike other ground-based gamma-ray observatories (e.g. H.E.S.S., VERITAS), which employ a large number of variables to separate the primary particles. In this work, we explore machine learning techniques (Boosted Decision Trees and Neural Networks) to identify the primary particles detected by HAWC. Our new gamma/hadron separation techniques were tested on data from the Crab nebula, the standard reference in Very High Energy astronomy, showing an improvement compared to the standard HAWC background rejection method.
}

\FullConference{37$^{\rm{th}}$ International Cosmic Ray Conference (ICRC 2021)\\
		July 12th -- 23rd, 2021\\
		Online -- Berlin, Germany}

\begin{document}
\maketitle

\section{Introduction}
	\paragraph*{}The beginning of Very High Energy (VHE, $>$ 100 GeV) gamma-ray astronomy could be said to be the TeV detection of the Crab Nebula (hereafter referred to simply as `the Crab') by the Whipple collaboration in 1989~\citep{weekes1989}. The challenge was removing the huge flux of cosmic rays, which lose any directional information because they are deflected by magnetic fields, while keeping most gamma rays, which represent a very small fraction of the total particles that impinge on Earth~\citep{bookVHE}. Therefore, ground-based detectors rely on gamma/hadron separation (hereafter referred to as `G/H separation') techniques to distinguish the showers that were initiated by a gamma ray from those initiated by charged particles; this is critical to improve the resolution and sensitivity of the instrument~\citep{probeCNNCTA}. 		
	\paragraph*{}Fegan~\citep{SumGHSepFegan} reviewed the evolution of the various efforts on G/H separation, starting with the application of a cut on Hillas variables (length, width, direction, etc) which result from a parametrization of the shower image; subsequently, a combination of these variables was employed to define a region where gamma rays are present; finally, machine learning techniques (MLT) have been employed to build or compute a region that is bounded by the data.
	\paragraph*{}MLTs are part of a new field in Computer Science and Statistics whose goal is to learn directly from data, creating powerful tools to build complex models for specific tasks, such as G/H separation. Nowadays, these kind of techniques are widely used in a variety of aspects of daily life, like machine translators. MLT are also used in high-energy astrophysics, by observatories like MAGIC and H.E.S.S., who have been using these methods to improve their G/H separation efforts~\citep{krause17,ohm09}. 
	\paragraph*{}In this work we use the two of the most popular MLTs: Neural networks (NNModel) and Boosted Decision Trees (BDTModel) to build a model to recognize and separate gamma-induced showers from hadron-induced ones. Since this kind of technique uses data to build these models, HAWC and its data will be described in Section \ref{sec:hawc}. Later, we show in Section \ref{sec:comparison} the comparison between the simulation and real data to check that they have the same behavior. This gives us confidence that we will obtain the comparable results using real data. Next, we describe the MLTs in Section \ref{sec:MLT} and their performance. We then provide a comparison with the official variables for G/H separation in Section \ref{sec:results}. Finally, in Section \ref{sec:conclu}, we discuss our results and plans for possible future improvements.
\section{HAWC}\label{sec:hawc}
	\paragraph*{}The High-Altitude Water Cherenkov (HAWC) gamma-ray observatory is located in the Sierra Negra volcano, in the state of Puebla, Mexico, 4,100 meter above sea level. HAWC is an array of 300 Water Cherenkov Detectors (WCD). Each WCD is instrumented with four photomultiplier tubes (PMTs) that detect the Cherenkov light produced by the secondary particles crossing the detector. The pulse of these PMTs is sent to the HAWC data acquisition system, where the charge and time of each `hit' is recorded. The next step involves the reconstruction of each event, using the saved information. The output of this process is a list of variables that contain the key characteristics of the cascade and specific information of the event~\citep{Abeysekara2017}.
	\paragraph*{}The HAWC data simulation uses several standard packages to reproduce the detection of one Crab transit with its background, that is, the gamma-ray and eight nuclei particles are simulated~\citep{Abeysekara2017}: First, Corsika is used to simulate the shower generated by the primary particles with an index of -2.00 (ultimately, each event is weighted to arrive at the right Crab index, -2.63); next, GEANT4~\citep{geant4reference} is used to simulate the secondary particles crossing through the WCD; next, the HAWC software generates the simulation of the PMT response; finally all this information is reconstructed in order to obtain the same list of variables as in the case of real data.
	\paragraph*{}Two variables from this list are currently used by the HAWC Collaboration for predicting the nature of the primary particle: Compactness and LIC. The method applies two simple cuts on each variable (in this work, we refer to this as the standard cut, SC)~\citep{Abeysekara2017}. The optimization cut depends on the number of PMTs active during the event and the estimated primary energy, so the data is split into two dimensional bins: the fraction of PMTs active during the event bin, $\mathcal{B}$ (see Table~\ref{Tab:fbin}),  and the Neural Network energy estimate bin, ebin (see Table~\ref{Tab:ebin})~\citep{energyestimatorpaper}.
	\begin{table}[]
		\begin{minipage}{.5\linewidth}
			\caption{{\small Fraction hit bins ($\mathcal{B}$).}}
			\centering
			\label{Tab:fbin}
			\scalebox{1.0}{
				\begin{tabular}{| c | c | }
					\hline
					$\mathcal{B}$ & Range (\%)  \\
					\hline
					0	& 	 4.4~-~~~6.7\\
					\hline
					1	& 	 6.7~-~~10.5\\
					\hline
					2	& 	10.5~-~~16.2\\
					\hline
					3	& 	16.2~-~~24.7\\
					\hline
					4	& 	24.7~-~~35.6\\
					\hline
					5	& 	35.6~-~~48.5\\
					\hline
					6	& 	48.5~-~~61.8\\
					\hline
					7	& 	61.8~-~~74.0\\
					\hline
					8	& 	74.0~-~~84.0\\
					\hline
					9	& 	84.0~-~100.0\\
					\hline
				\end{tabular}
			}
		\end{minipage}%
		\begin{minipage}{.5\linewidth}
			\centering
			\caption{{\small The energy bins ($ebin$).}}
			\label{Tab:ebin}
			\scalebox{0.85}{
				\begin{tabular}{| c  | c | }
					\hline
					ebin & Range ($log(${\bf \^E}$/GeV)$) \\
					\hline
					a & 2.50 - 2.75\\
					\hline
					b & 2.75 - 3.00\\
					\hline
					c & 3.00 - 3.25\\
					\hline
					d & 3.25 - 3.50\\
					\hline
					e & 3.50 - 3.75\\
					\hline
					f & 3.75 - 4.00\\
					\hline
					g & 4.00 - 4.25\\
					\hline
					h & 4.25 - 4.50\\
					\hline
					i & 4.50 - 4.75\\
					\hline
					j & 4.75 - 5.00\\
					\hline
					k & 5.00 - 5.25\\
					\hline
					l & 5.25 - 5.50\\
					\hline
				\end{tabular}
			}
		\end{minipage} 
	\end{table}
\section{Data/simulation comparison}\label{sec:comparison}
	\paragraph*{}We explore all variables in the reconstruction list. However not all of them have a good performance and are suitable to be used for G/H separation. Only five of these variables were chosen to feed into the MTL: LIC, PINCness, disMax, LDFChi2 and LDFAmp. These five variable focus on the main differences between gamma-induced and hadron-induced showers: the number of muons in the shower or the spread of the shower in the experiment. Before starting the training \& verification stage, we checked that these variables have the same behavior in the simulation and the real data. 
	\doublefig{0.8}{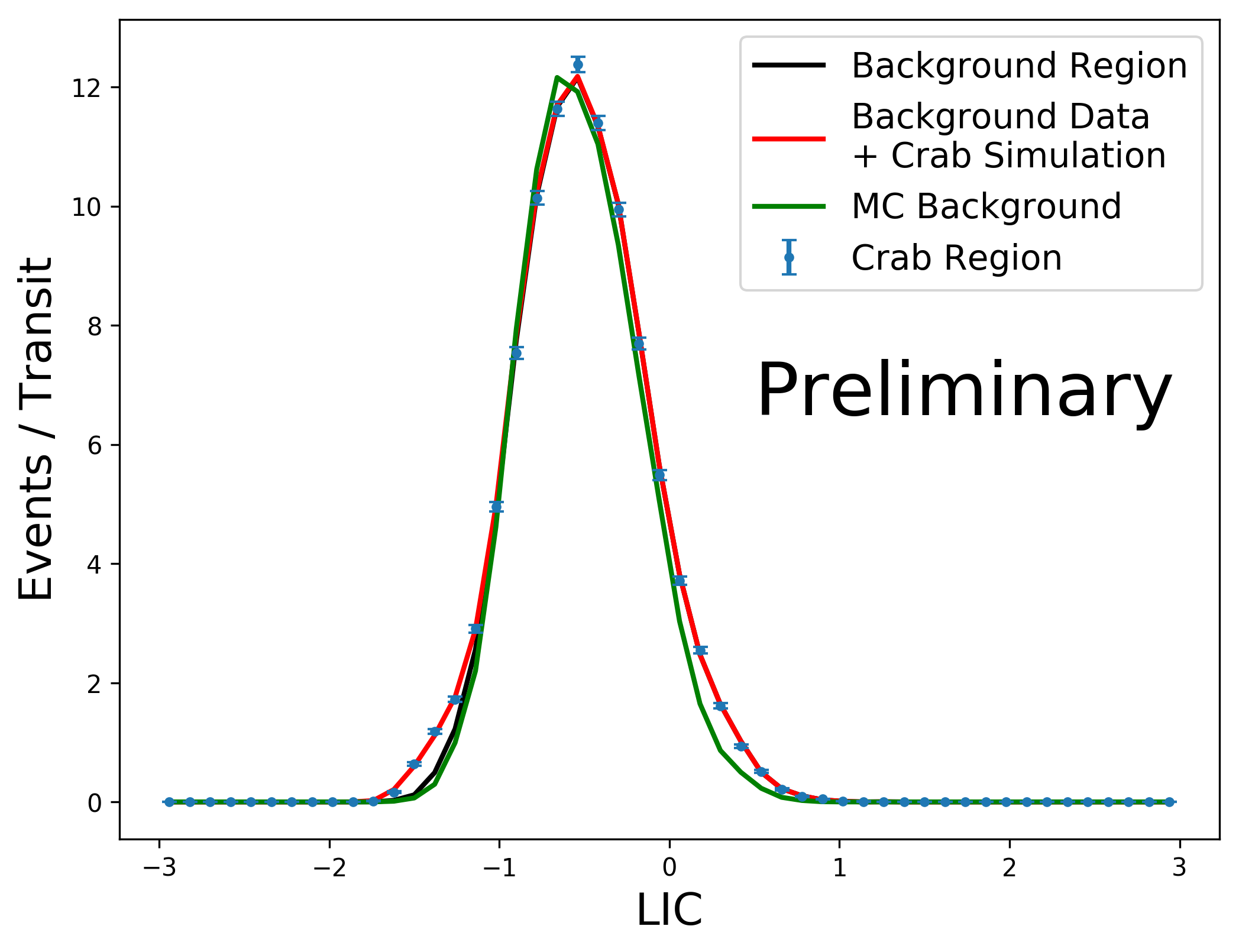}{Background comparison}{Fig:LIC06bkg}{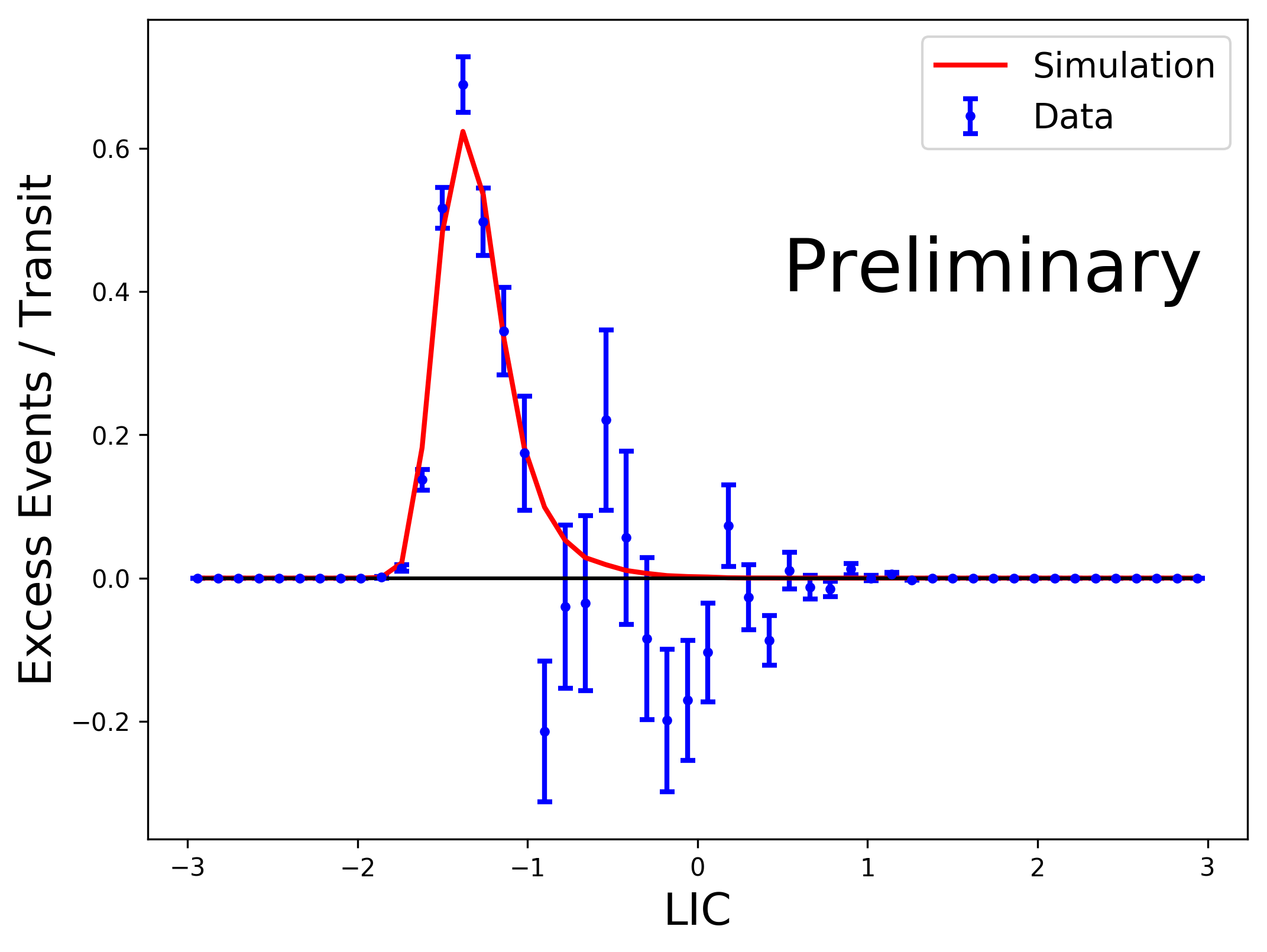}{Signal comparison}{Fig:LIC06Sig}{Comparison between Crab simulated and real data, focusing on the LIC variable, for $\mathcal{B}$ 6. {\bf a)} Comparison of the background: the black line shows the off-Crab data and the green line shows simulated background, both identical in shape. In addition, we compare the signal: blue points show the Crab data (signal plus background), while the red line is the MC simulation of signal plus background (off-Crab data), both showing the same behavior.  {\bf b)} shows more clearly the signal comparison: the red line represents the MC simulated signal and the blue line shows actual Crab data}{Fig:LIC06}
	\paragraph*{}We used Crab simulations and real data around the Crab to carry out our data/simulation comparison. The simulation uses a power law spectrum with an index of 2.63, and we generated two simulation data sets: first one contains gamma-ray events (referred to as the MC signal simulation) while the second includes nuclei particles (referred to as MC background simulation). On the other hand, a real background data set is easy to obtain because approximately 99.9\% of all primary particles generating HAWC events are background, but the signal is more difficult. Thus, a long period (837 live days\footnote{From June 2015 to December 2017}) and a subtraction technique are used to estimate the Crab signal\footnote{Basically this technique involves taking the events from the Crab (signal plus background) and subtracting the off-Crab events (only background). The result obtained is the pure signal from the Crab.}\citep{Abeysekara2017,ohm09}.
	\paragraph*{}Figure \ref{Fig:LIC06} and \ref{Fig:PINC05} show the results  of two powerful G/H separation variables for different values of $\mathcal{B}$: LIC and PINC. In the figures, panel a) shows the background comparison and signal plus background comparison, at the same time. The simulation follows the real data, that is, the background histogram has the same behavior, maybe in some bins with a small offset, and the prediction of the signal plus background (red line) follows the shape of the blue point obtained from the Crab. In the b) panels of the Figures, we show the Signal comparison, that is, the histogram of the MC signal simulation and the Crab signal using the subtraction technique. Both histograms look similar, despite certain fluctuations that could be caused by a lack of events in a region where we do not expect many gamma rays.
	\doublefig{0.8}{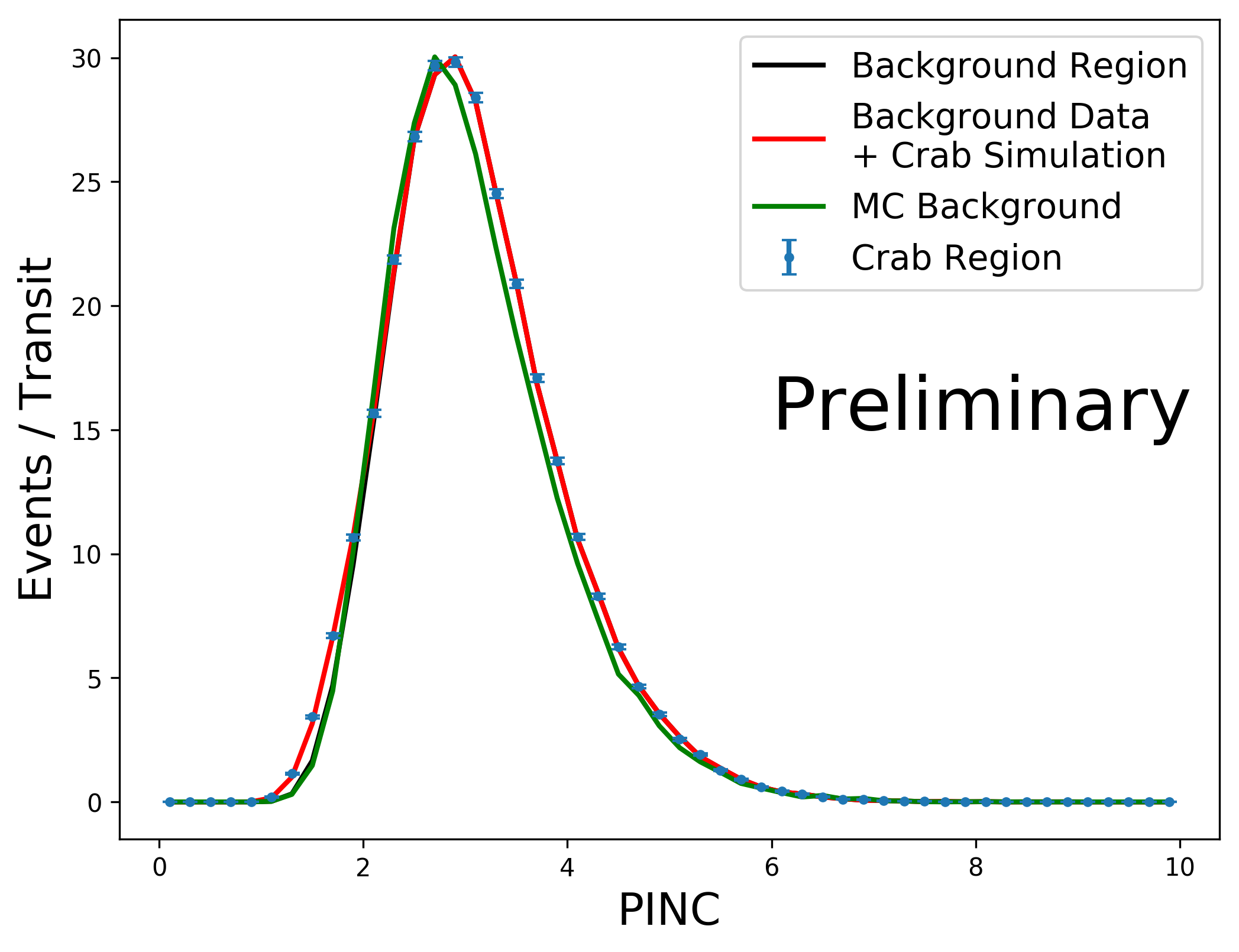}{Background comparison}{Fig:PINC05bkg}{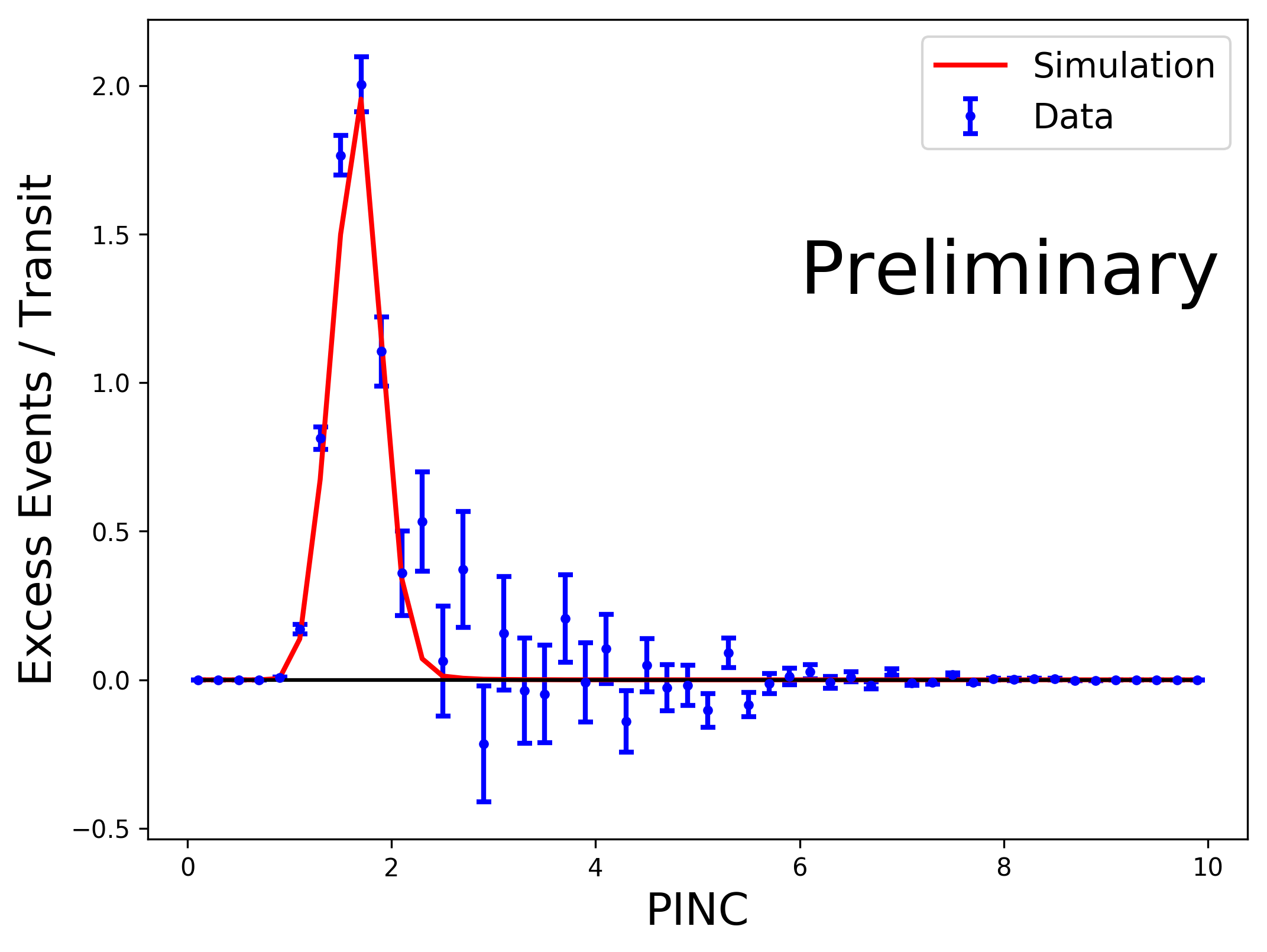}{Signal comparison}{Fig:PINC06Sig}{Same as Figure~\ref{Fig:LIC06} but here using the PINC variable and $\mathcal{B}$ 5}{Fig:PINC05}
\section{G/H separation using MLT}\label{sec:MLT}
	\paragraph*{}The NNModel and BDTModel are fed with seven variables: the five powerful G/H separation variables mentioned above, the fraction of PMTs active during the event, and the Neural Network energy estimator. The last two variables were added because the value of the other five variables depends on them. Three models were trained for each MLT, operating on specific $\mathcal{B}$ (0-2, 3-5 and 6-9). We used the toolkit for multivariate data analysis (TMVA) in ROOT~\citep{tmvareference} to implement the NNModels, and the Gradient Boosted Decision python package~\citep{chen2016xgboost} for our BDTModel. In the Training \& Verification stage, we used only 50 \% of the simulation, leaving the rest to use in the Testing stage (see the next section). 
\section{Results}\label{sec:results}
	\paragraph*{}The performance of the MLT was tested using known events, that is, only the 50\% of the simulated. The goal of the G/H separation is to remove the maximum number of background events and retain the most signal events. Thus, the events of interest are the gamma events, so most of them, if not all, should be included in the source analysis and in order to not overshape the source, the hadron events must be removed. Figure \ref{Fig:sigeff} shows the gamma efficiency and the MLTs have better results, with their performance depending on $\mathcal{B}$. However, the hadron efficiency (Figure \ref{Fig:bkgeff}) of the SC is better than the MLTs after $\mathcal{B}$ 7; below this bin, the BDTModel have quite similar results to the SC, while the NNModel is the worst between $\mathcal{B}$ 3 to 6. 
	\esqfig{0.8}{0.4}{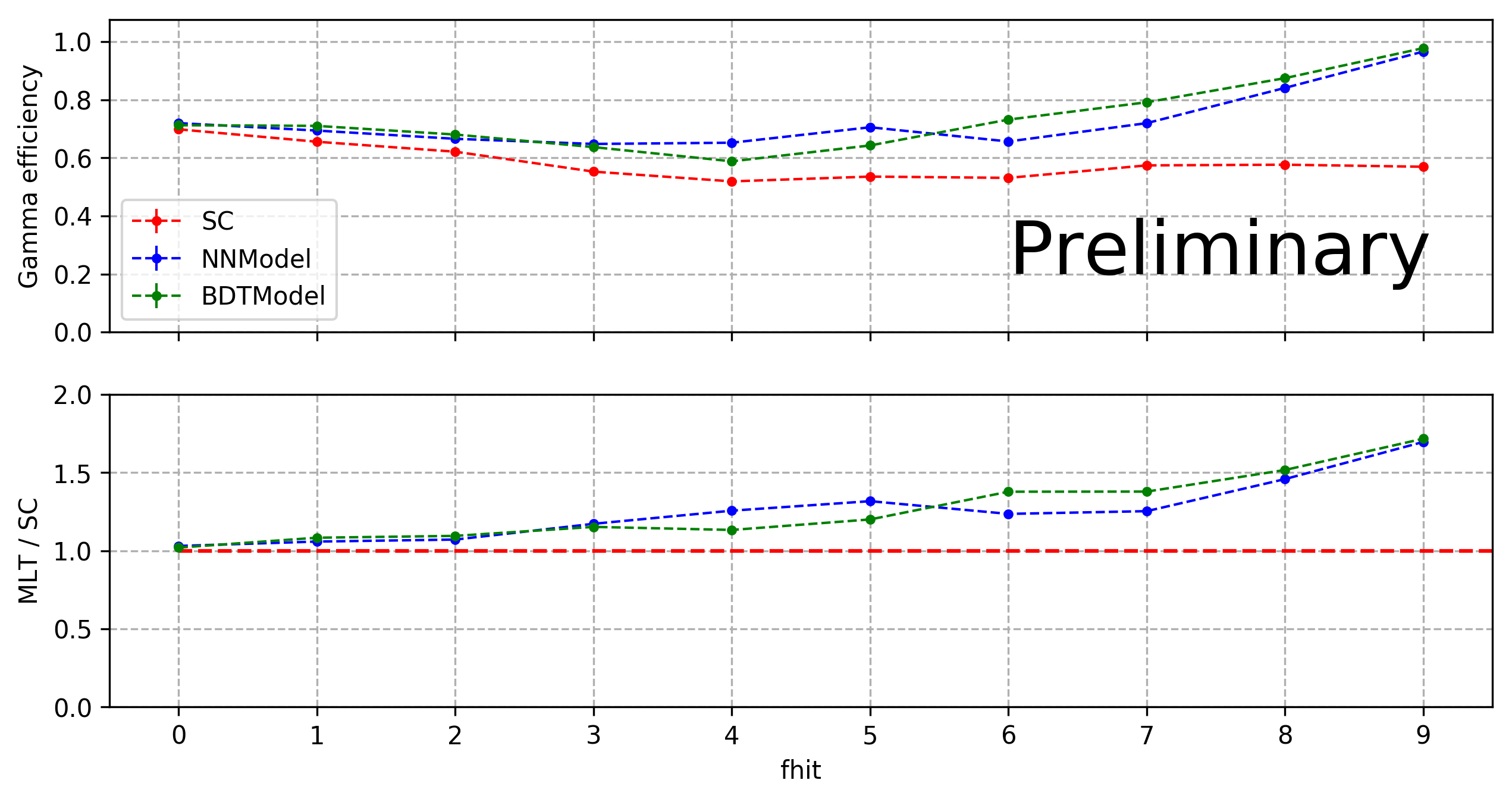}{Top panel shows the gamma efficiency of the three G/H separation techniques, while the bottom panel shows the ratio of the MLTs over the SC}{Fig:sigeff}	
	\esqfig{0.8}{0.4}{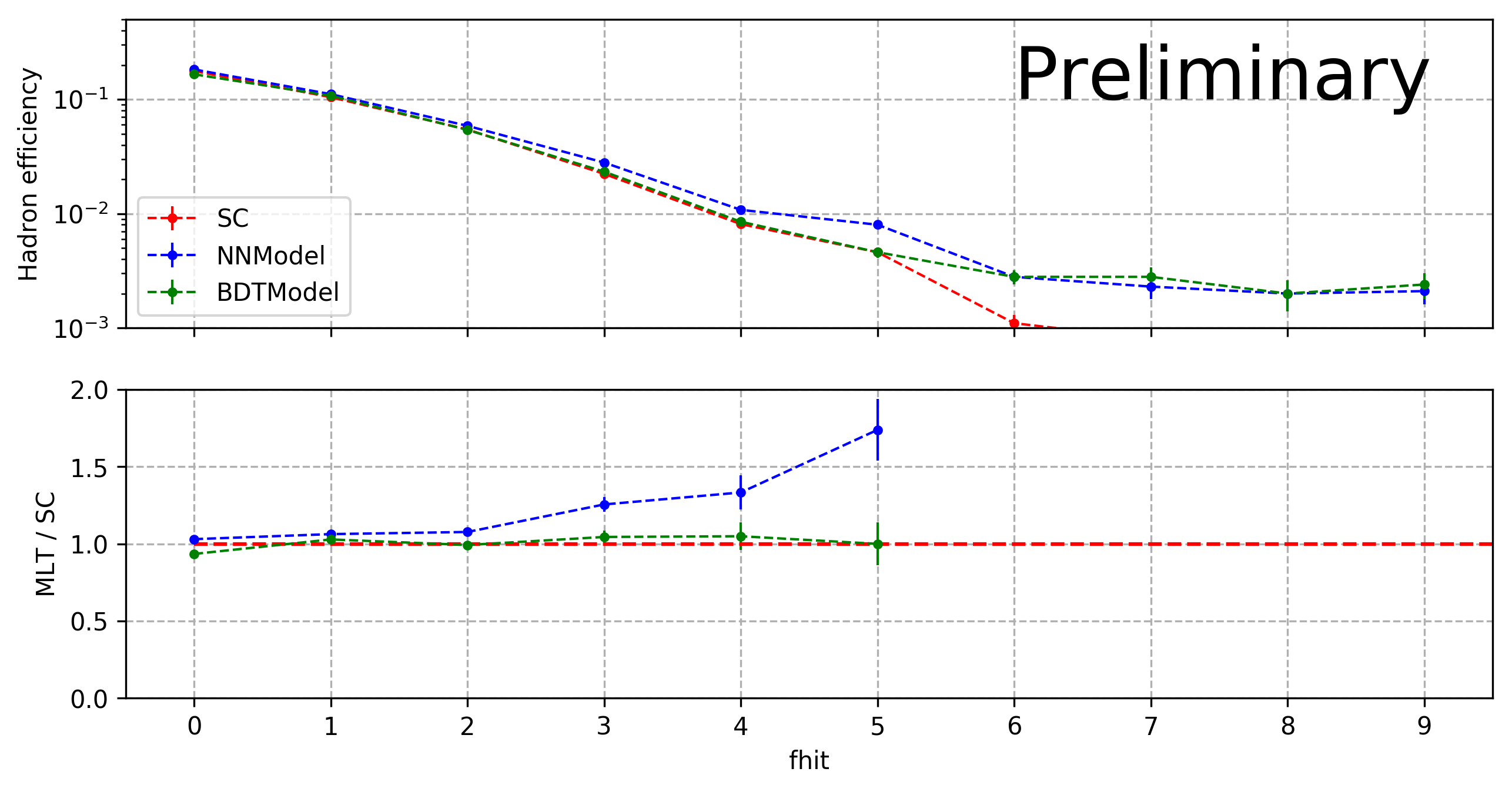}{Same as Figure \ref{Fig:sigeff} but here is the hadron efficiency. After $\mathcal{B}$ 6, the SC have good hadron rejection (0.01\%) that the MLT/SC ratio value have greater than 2}{Fig:bkgeff}
	\paragraph*{}The most robust test of the MLTs comes from applying them to the real data, thus we use 837 live days of Crab data. Figure \ref{Fig:Crabfhit} shows the significance of the Crab using the three G/H separation techniques, with the best model being the BDTModel. The NNModel is better than the SC in some bins, but not all, while the performance of the NNModel and SC are similar if ebin is used (Figure \ref{Fig:Crabfhit}), and the BDTModel shows the most improvement in the middle energy bands (3.25 to 4.25).
	\esqfig{0.8}{0.4}{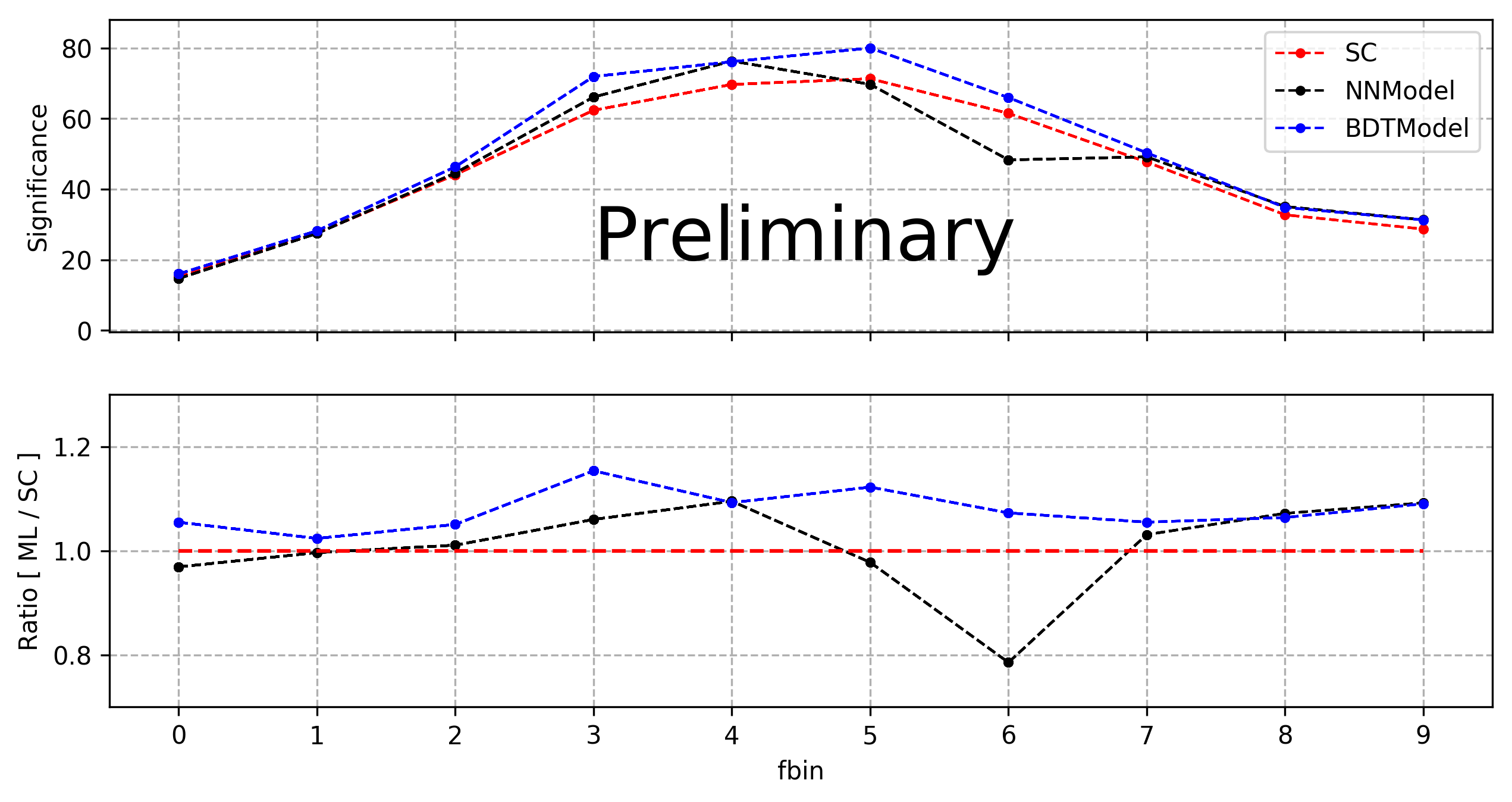}{Top panel shows the Crab significance using each G/H separation technique as a function of $\mathcal{B}$. Bottom panel shows the ratio of each MLT over the SC}{Fig:Crabfhit}
	\esqfig{0.8}{0.4}{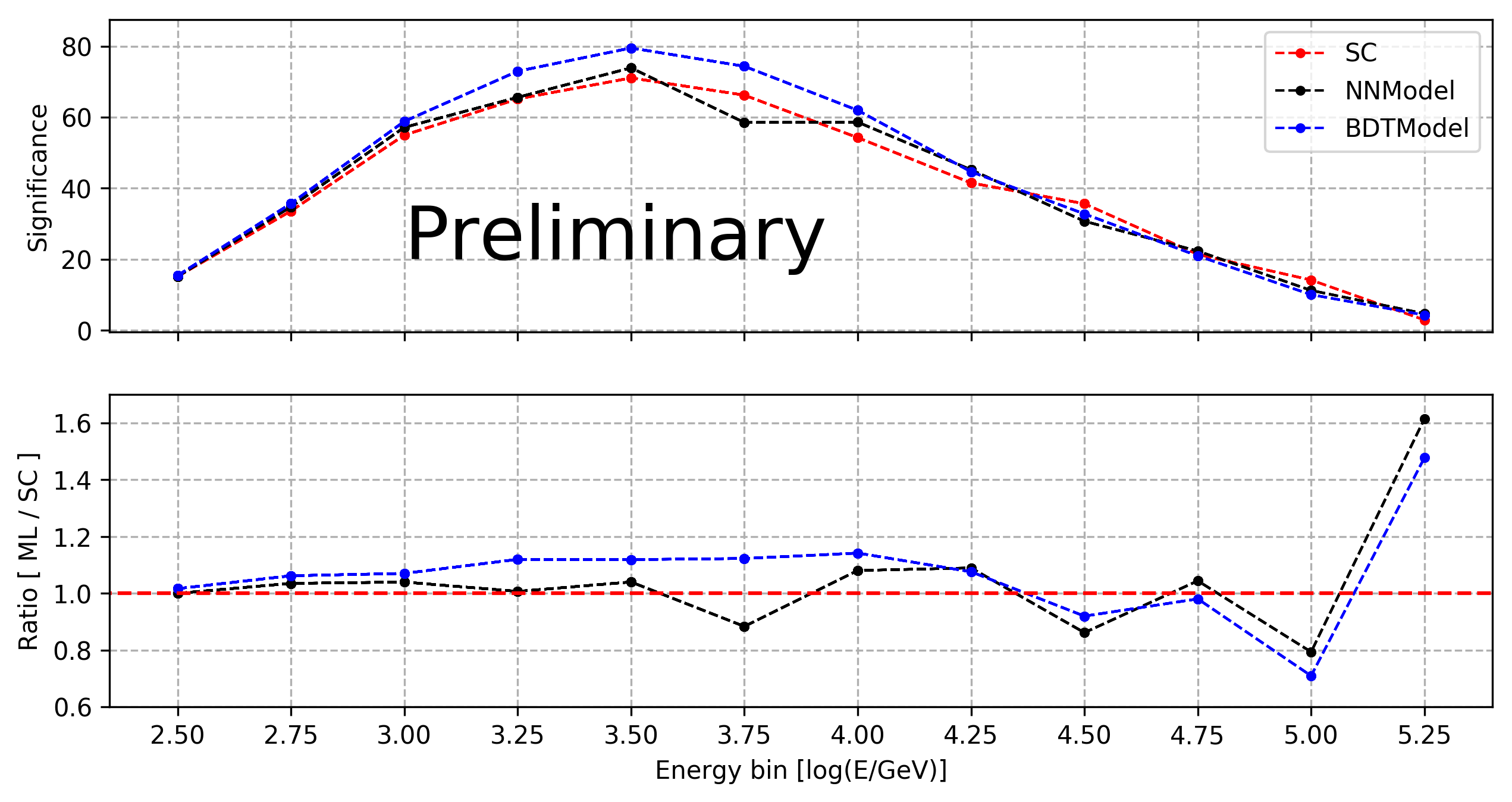}{Same as Figure \ref{Fig:Crabfhit} but here report for each ebin}{Fig:Crabebin}
\section{Conclusion and Discussion}\label{sec:conclu}
	\paragraph*{} In this work, we explore the use of MLTs for building complex models for G/H separation with the HAWC observatory. A NNModel and BDTModel were trained using 50\% of a MC simulation, with the rest being used for testing. The MLTs show a good performance in classifying gamma events but a worse result compared to the SC. Both MLTs are applied to obtain the significance at the Crab. The best G/H separation is the BDTModel, however there are some ebin that NNModel or SC have better results. The NNModel shows similar results to the SC. 
	\paragraph*{}The results of this work has reported a potential to improve results on the G/H separation task that MLT can be used to build complex models and obtain better result of both HAWC and other WCD array in the future. 
	\paragraph*{}The MLT is a wide area of research that can provide a variety of models for improving on the G/H separation task. For example, the work of Lyard et. at.~\citep{probeCNNCTA} reported the comparison between a BDTModel and a convolution neural network (CCN) using Cherenkov Telescope of the CTA Collaboration, where the CNN show an improvement over the BDTModel after 1TeV. Another effort is the work of Watson et at. \citep{ianicrc2021} that employed a CCN to separate gamma rays from background in HAWC. 
\acknowledgments{
We acknowledge the support from: the US National Science Foundation (NSF); the US Department of Energy Office of High-Energy Physics; the Laboratory Directed Research and Development (LDRD) program of Los Alamos National Laboratory; Consejo Nacional de Ciencia y Tecnolog\'ia (CONACyT), M\'exico, grants 271051, 232656, 260378, 179588, 254964, 258865, 243290, 132197, A1-S-46288, A1-S-22784, c\'atedras 873, 1563, 341, 323, Red HAWC, M\'exico; DGAPA-UNAM grants IG101320, IN111716-3, IN111419, IA102019, IN110621, IN110521; VIEP-BUAP; PIFI 2012, 2013, PROFOCIE 2014, 2015; the University of Wisconsin Alumni Research Foundation; the Institute of Geophysics, Planetary Physics, and Signatures at Los Alamos National Laboratory; Polish Science Centre grant, DEC-2017/27/B/ST9/02272; Coordinaci\'on de la Investigaci\'on Cient\'ifica de la Universidad Michoacana; Royal Society - Newton Advanced Fellowship 180385; Generalitat Valenciana, grant CIDEGENT/2018/034; Chulalongkorn University’s CUniverse (CUAASC) grant; Coordinaci\'on General Acad\'emica e Innovaci\'on (CGAI-UdeG), PRODEP-SEP UDG-CA-499; Institute of Cosmic Ray Research (ICRR), University of Tokyo, H.F. acknowledges support by NASA under award number 80GSFC21M0002. We also acknowledge the significant contributions over many years of Stefan Westerhoff, Gaurang Yodh and Arnulfo Zepeda Dominguez, all deceased members of the HAWC collaboration. Thanks to Scott Delay, Luciano D\'iaz and Eduardo Murrieta for technical support.
}		

\bibliographystyle{JHEP} 
\bibliography{biblio}

\clearpage
\section*{Full Authors List: \Coll\ Collaboration}
\scriptsize
\noindent
A.U. Abeysekara$^{48}$,
A. Albert$^{21}$,
R. Alfaro$^{14}$,
C. Alvarez$^{41}$,
J.D. Álvarez$^{40}$,
J.R. Angeles Camacho$^{14}$,
J.C. Arteaga-Velázquez$^{40}$,
K. P. Arunbabu$^{17}$,
D. Avila Rojas$^{14}$,
H.A. Ayala Solares$^{28}$
% ...

\noindent
% ...
$^{14}$Instituto de Física, Universidad Nacional Autónoma de México, Ciudad de México, México,
% ...
$^{17}$Instituto de Geofísica, Universidad Nacional Autónoma de México, Ciudad de México, México,
% ...
$^{21}$Physics Division, Los Alamos National Laboratory, Los Alamos, NM, USA,
% ...
$^{28}$Department of Physics, Pennsylvania State University, University Park, PA, USA,
% ...
$^{40}$Instituto de Física y Matemáticas, Universidad Michoacana de San Nicolás de Hidalgo, Morelia, Michoacán, México,
$^{41}$FCFM-MCTP, Universidad Autónoma de Chiapas, Tuxtla Gutiérrez, Chiapas, México,
% ...
$^{48}$Department of Physics and Astronomy, University of Utah, Salt Lake City, UT, USA,
% ...

\end{document}